\documentclass[journal=mamobx,manuscript=letter,layout=traditional]{achemso}

\usepackage[version=3]{mhchem} % Formula subscripts using \ce{}
\usepackage{graphicx,amssymb,amstext,amsmath} % Include figure files
\usepackage{color}
\usepackage{url}

\author{Richard Matthews}

\affiliation[]
{Faculty of Physics, University of Vienna, Boltzmanngasse 5, A-1090 Vienna, Austria}

\email{richard.matthews@univie.ac.at}

\author{Ard A. Louis}

\affiliation[]
{Rudolf Peierls Centre for Theoretical Physics, 1 Keble Road, Oxford 0X1 3NP, England}

\author{Christos N. Likos}

\affiliation[]
{Faculty of Physics, University of Vienna, Boltzmanngasse 5, A-1090 Vienna, Austria}

\title[]
  {Effect of bending rigidity on the knotting of a polymer under tension \footnote{This document is the unedited author's version of a Submitted Work that
was subsequently accepted for publication in ACS Macro Letters, copyright \copyright
American Chemical Society after peer review. To access the final edited
and published work, see \url{http://pubs.acs.org/articlesonrequest/AOR-DTxiaBcAyZmvaibHpdvq}}}

\begin{document}
%%%%%%%%%%%%%%%%%%%%%%%%%%%%%%%%%%%%%%%%%%%%%%%%%%%%%%%%%%%%%%%%%%%%%
%% The manuscript does not need to include \maketitle, which is
%% executed automatically.  The document should begin with an
%% abstract, if appropriate.  If one is given and should not be, the
%% contents will be gobbled.
%%%%%%%%%%%%%%%%%%%%%%%%%%%%%%%%%%%%%%%%%%%%%%%%%%%%%%%%%%%%%%%%%%%%%
\begin{abstract}
A coarse-grained computational model is used to investigate how the bending rigidity of a polymer under tension affects the formation of a trefoil knot. Thermodynamic integration techniques are applied to demonstrate that the free-energy cost of forming a knot has a minimum at non-zero bending rigidity. The position of the minimum exhibits a power-law dependence on the applied tension. For knotted polymers with non-uniform  bending rigidity, the knots preferentially localize in the region with a bending rigidity that minimizes the free-energy.
\end{abstract}


Type II topoisomerases are enzymes that may knot or unknot DNA by introducing a transient break in both strands of one DNA duplex and passing a second duplex through it. One of their key biological functions is to regulate the level of knotting in the genome~\cite{rybenkov}. Type II topoisomerases tend to act preferentially on certain sequences in DNA~\cite{burden}. There is evidence that sites that are more frequently cleaved tend to be located in or next to parts of the genome called scaffold associated regions or matrix attachment regions~\cite{adachi,razin,masliah}, which are typically several hundred base pairs long~\cite{adachi} and rich in adenine (A) and thymine (T), two of the nucleotides in DNA. Further, a specific sequence evolved $in$ $vitro$, which was preferentially cleaved by a certain type II topoisomerase, was highly AT-rich~\cite{burden}.

It is believed that AT-rich sequences are more flexible than random ones~\cite{masliah,hogan,okonogi,scipioni}. For example, the work of Okonogi {\em et al.}~\cite{okonogi}  suggests that a sequence of AT repeats is about 20\% more flexible than a random sequence. An earlier study  suggested that such an AT rich sequence can have a persistence length less than half that of a GC rich sequence~\cite{hogan}. Scipioni $et$ $al.$~\cite{scipioni}  used scanning force microscopy to observe a correlation between AT-rich parts of a DNA fragment and flexibility. Further, Masilah $et$ $al.$~\cite{masliah} found that there is a preferentially large opening of the base-pairs immediately adjacent to a preferentially cleaved site. This opening was found to be dependent on the sequence context. Opening of base-pairs (bubble formation) can lead to greatly increased local flexibility~\cite{ramstein}. Very high flexibility at the topoisomerase II cleavage sites is probably necessary because the enzyme enforces a large bend in DNA when it binds to it~\cite{dong}.

An intriguing question arises as to whether the correlation between the positions of cleavage sites and DNA flexibility could be important in the regulation of knotting. For example, could the variation of bending stiffness help to localize knots near cleavage sites, thus expediting their removal? Here we make a first step towards understanding these issues by using a simple bead-spring polymer model to investigate how the free energy cost of forming a knot, $\Delta F_{knotting}$, varies with polymer bending stiffness and how this influences the position of a knot within a polymer of non-uniform flexibility. In this work, we simulate only the trefoil knot, $3_1$~\cite{livingston}, but our general arguments do not depend on the particular topology. Previous work~\cite{liu3} on how the action of type II topoisomerase may be guided by bent geometries of DNA has been performed, but variable bending stiffness was not considered.

The case of polymers under tension is biologically relevant because the action of enzymes during processes such as transcription applies forces to DNA~\cite{bustamante2003,liu3}. In general, for polymers in a good solvent with bending stiffness, $A$, under tension, $\tau$, there are three main contributions to $\Delta F_{knotting}$: the reduction in entropy due the self-confinement of the polymer in the knotted region; the increase in bending energy due to the curvature enforced by the knot; and the work done against the tension in reducing the extension of the polymer, necessary to give free length for knot formation.

We consider how $\Delta F_{knotting}$ varies with $A$ for fixed $\tau$. We identify two length scales: that associated with the bending stiffness, $l_{A} \sim A/(k_B T)$, and that associated with the size of the knotted region, $l_{knot}(A)$, which depends on $A$. When $l_{A} \ll l_{knot}(A)$ the main effect of increasing $A$ will be to decrease the entropic cost of knotting and $\Delta F_{knotting}$ will decrease with $A$. Previous work  on fully flexible chains ($A = 0$), has found knots to be weakly localized~\cite{farago,marcone,mansfield2010}, $N_{knot} \sim N^t$, where $N_{knot}$ is the number of monomers in the knot, $N$ the total number in the polymer, and $0 < t < 1$~\cite{farago}. By applying scaling arguments based on the blob picture to interpret the results of simulations of polymers under tension, Farago $et$ $al.$~\cite{farago}, estimated $t = 0.4\pm0.1$. A later study used two methods, including one based on closing subsections of the polymer and calculating a knot invariant, to find $t \simeq 0.75$~\cite{marcone}. The discrepancy between the two estimates may be attributed to the relatively short polymers used in the earlier work~\cite{marcone}. Knot localization has been observed experimentally~\cite{ercolini} but is found to disappear with confinement.~\cite{tubiana}. A free energy calculation for an open, linear polymer found no evidence of a metastable knot size.~\cite{zheng2010}

\begin{figure}[htb!]
\begin{center}
\includegraphics[scale=0.55]{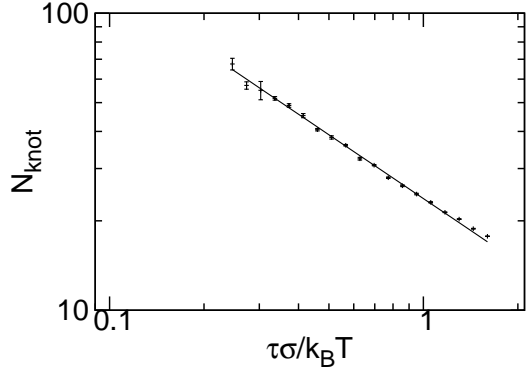}
\caption{\label{fig:ks_tens} Variation of the number of beads forming the knot, $N_{knot}$ with tension, $\tau$ for $N = 256$ bead flexible polymers. The solid line is a fit to the data with slope $-0.71\pm0.01$. Errorbars were estimated by performing three independent repeats of the simulations}
\end{center}
\end{figure}

In the flexible regime, a polymer under tension will form a linear series of blobs of $N_b \sim (k_B T/\tau)^{1/\nu}$ monomers each, where $\nu \approx 3/5$~\cite{deGennes_scpp}. The series of blobs cannot be knotted and so the knot resides within one blob. Treating this blob as an independent polymer, we expect $l_{knot}$ to be determined by the entropic localization of the knot and the number of monomers participating to the knot to scale, accordingly, as $N_{knot} \sim (k_BT/\tau)^{t/\nu}$. By employing the simulation techniques and knot-identification algorithm to be presented shortly, we have determined the dependence of $N_{knot}$ on $\tau$ for a flexible polymer of $N = 256$ beads of size $\sigma$ each. The results in~\ref{fig:ks_tens} indeed show a power-law dependence. By fitting to this data, we estimate that $t = 0.43 \pm 0.01$, which is consistent with the value found by Farago $et$ $al.$~\cite{farago}, as expected given the relatively short chains used. Concomitantly, the knot size in {\it fully flexible chains} scales as $l_{knot}(0) \sim N_{knot}^{\nu} \sim (k_BT/\tau)^t$.

For $l_{A} \gg l_{knot}(A)$ the size of the knot will be dominated by the interplay of bending energy and tension and $\Delta F_{knotting}$ will increase with $A$. We therefore expect a minimum of $\Delta F_{knotting}(A)$ at a value of $A$ determined by $l_{A} \approx l_{knot}(A)$. As the dependence of $l_{knot}(A)$ on $\tau$ is not known, we replace $l_{knot}(A)$ with $l_{knot}(0)$ to find what the likely form of the dependence of the bending stiffness for which the free energy cost is minimal, $A_{min}$, on $\tau$ is. Using the results obtained above, a power-law dependence is obtained:
\begin{equation}\label{eq:scalea}
A_{min} \sim \tau^{-t}.
\end{equation} 

Of course, the replacement of $l_{knot}(A)$ with $l_{knot}(0)$ in the relationship $l_{A} \approx l_{knot}(A)$ is an approximation which is expected to break down precisely in the region of validity of this equality. On the other hand, a power-law dependence $N_{knot} \sim N^{t_A}$ is a reasonable assumption also for the case $A \neq 0$, thus we anticipate a relationship of the form of~\ref{eq:scalea} to hold also for $A \neq 0$, albeit with some exponent $t_A \ne t$.

For very large values of $A$, we expect the knot to form a single loop with the all crossings close to each other~\cite{gallotti}. Assuming the thickness of the polymer is small compared to the loop, we expect $\Delta F_{knotting}$ to be approximately given by~\cite{gallotti}
\begin{equation}\label{eq:D_F_stiff}
\Delta F_{knotting} = \sqrt{8 \pi^2 A \tau}.
\end{equation}
For lower $A$, the form of $\Delta F_{knotting}$ may not be so easily deduced. At the crossover this is particularly difficult because here we expect the bending length and self-confinement length to be approximately equal. For this case, a scaling form of the confinement free energy is not available~\cite{chen2006}.

We next study the consequences of these predictions with computer simulations. In what follows, we first outline the technical details of our approach, we then present results on $\Delta F_{knotting}$, before investigating the positional probability distribution of knots in polymers of non-uniform flexibility. We primarily simulate single chains of $N = 256$ beads of size $\sigma$ in a simulation box of volume $V = 2.048 \times 10^5 \sigma^3$ with periodic boundaries: unless otherwise stated, all results are for these parameters. The polymers are connected to themselves across the periodic boundaries in the $x$-direction. A constant tension is simulated by including in the potential a term proportional to the $x$-length of the box, $L_{x}$ and  allowing $L_{x}$ to vary. The advantage of this approach is that there are no free ends so that, as long as chain crossings are prevented, unknotting will never occur.

The simulation of the polymer is carried through for the following interaction potential:
\begin{eqnarray}\label{potential}
\nonumber V(\{{\bf r}_i\})= &-&\sum_{i}\kappa_{i} \left( \mathbf{\hat{r}}_{i-1,i} \cdot  \mathbf{\hat{r}}_{i,i+1} \right) - \tau L_{x}
\\\nonumber&-& \frac{k R_0^2}{2}\sum_{i}\ln\left[1-\left(\frac{r_{i,i+1}}{R_0}\right)^2\right]
\\\nonumber&+&  \sum_{j>i}\sum_{i}H\left[ 2^{\frac{1}{6}}\sigma - r_{i,j}\right] 
\\&\times& 4\epsilon\left[\left(\frac{\sigma}{r_{i,j}}\right)^{12}-\left(\frac{\sigma}{r_{i,j}}\right)^{6}+\frac{1}{4}\right],
\end{eqnarray} 
where $\mathbf{r}_{i,j} = \mathbf{r}_{j} - \mathbf{r}_{i}$, is the vector from bead $i$ to bead $j$, located at position vectors ${\bf r}_i$ and ${\bf r}_j$, respectively, whereas  $\mathbf{\hat r}_{i,j}$ denotes a unit vector. The first term sets the bending stiffness, which may be varied along the chain using the parameter $\kappa_{i}$, giving a bending stiffness of $A = \kappa_{i} \sigma$ for the $i$th bead. The second term applies a tension, $\tau$. The third and fourth terms are spring and excluded volume terms respectively, $H$ is the Heaviside step function which truncates the Lennard-Jones potential to be purely repulsive. We choose $\epsilon = k_BT$, $k = 30 k_BT / \sigma^2$ and $R_0 = 1.5\sigma$, which prevents the chain from crossing itself and so conserves topology.

We simulate using a Monte Carlo (MC) algorithm~\cite{frenkel}, which comprises two types of moves. To simulate a given tension, moves that attempt to change $L_{x}$, whilst rescaling $L_{y}$ and $L_{z}$ to keep $V$ fixed and also applying a corresponding transformation to all particle coordinates, are included. Displacements of the polymer beads are made using the Hybrid MC method~\cite{mehlig}, where trial states are generated using Molecular Dynamics (MD). During the MD trajectories, $L_{x}$ is fixed, the tension term is not included in the Hamiltonian used to calculate the forces. Collective motions of the polymer beads are more easily captured in this way than by local, single bead moves.

To calculate $\Delta F_{knotting}$ for a given tension, $\tau$, we simulate systems with all $\kappa_{i}$ set to the same value, $\kappa$. We simulate two sets of systems, one with linear topology and one with knotted polymers. The systems within one set span a range of rigidities from $\kappa = 0$ up to the desired value. For each of those values, we calculate the average $\left< \frac{\partial V}{\partial \kappa} \right>$. By numerically integrating $\left< \frac{\partial V}{\partial \kappa} \right>$ from $\kappa = 0$, we obtain the relative free energy as a function of $\kappa$~\cite{frenkel}, $\Delta F_{\alpha}(\kappa) = F_{\alpha} (\kappa) - F_{\alpha}(0)$, where $\alpha$ stands for either `knot' or `linear'. To fully determine $\Delta F_{knotting}$ we would need to perform an integration between unknotted and knotted states. However, since we are interested in the relative cost of knotting for different bending stiffnesses, we simply calculate $\Delta F_{knotting}(\kappa) - \Delta F_{knotting}(0) = \Delta F_{knot} (\kappa) - \Delta F_{linear} (\kappa)$ instead.

To improve the efficiency of our calculation of $\Delta F_{knotting}(\kappa) - \Delta F_{knotting}(0)$ we implemented the most computationally intensive part of our simulation algorithm on a GPU using CUDA, which allows for a high degree of parallelism but is restrictive in terms of the homogeneity of the parallel calculations~\cite{vanMeel2008}. Whilst a standard local-move MC algorithm would be difficult to implement on a GPU~\cite{vanMeel2008}, the most time-consuming part of our algorithm is calculating the MD trajectories to produce trial states for the Hybrid MC. The MD integration may be straightforwardly performed on a GPU. We simulate all systems for a given $\tau$ and topology in parallel, performing force calculations and integration steps on the GPU. As a simple alternative to a cell-list we reduce the number of pair separations calculated by exploiting the connectivity of the polymer, which guarantees the maximum separation of two beads within a section: by comparing the center of mass positions of two sections we can determine whether beads within them may interact. Random number generation and other MC moves were performed on the CPU. To help reduce correlation times we added parallel tempering~\cite{frenkel} swaps between systems with different $\kappa$.

\begin{figure}[htb!]
\begin{center}
\includegraphics[scale=0.25]{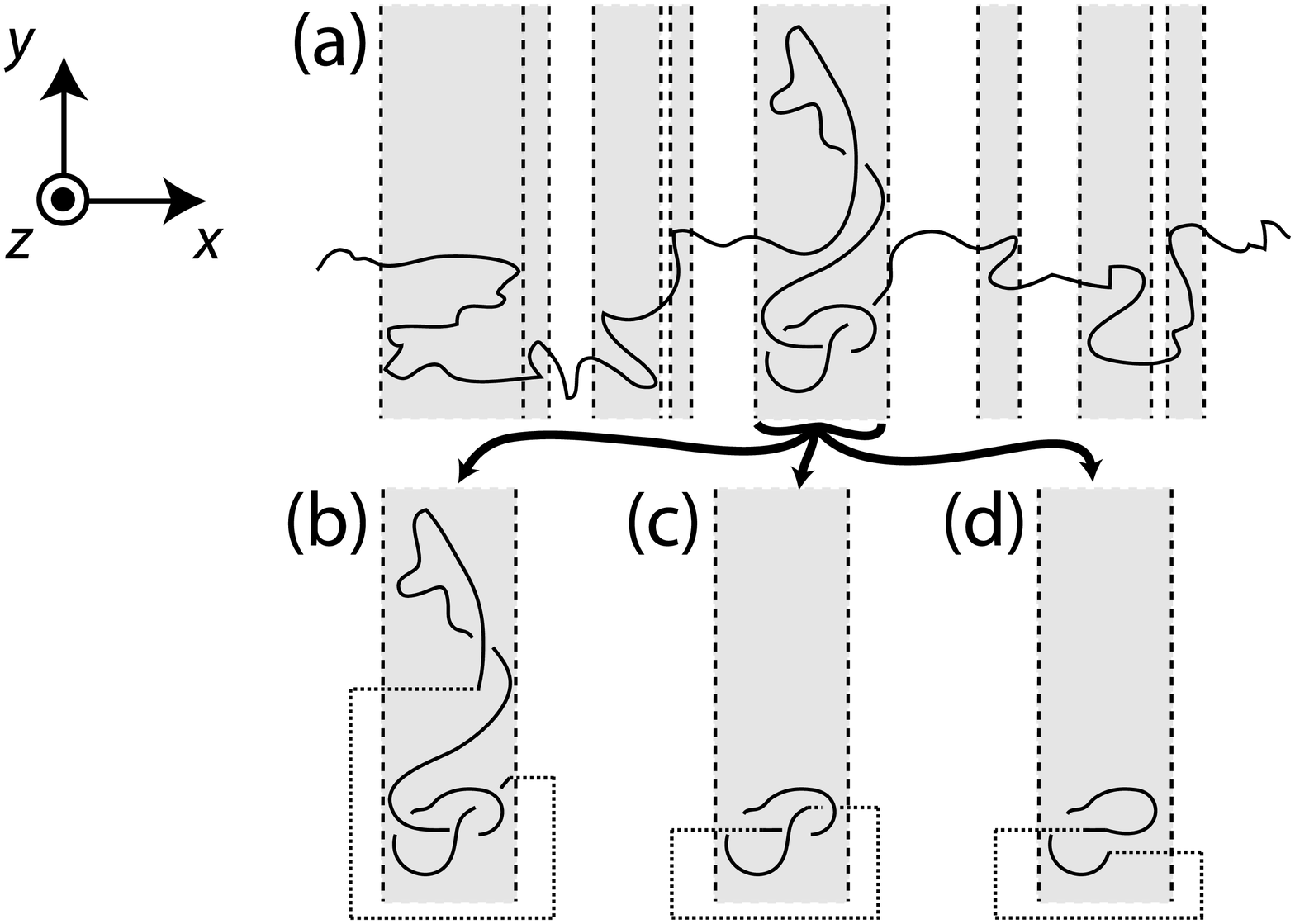}
\caption{\label{fig:knot_find_var_flex_tens} Schematic depiction of the knot-finding process. (a) The polymer is divided into sections by finding points along its contour -- indicated by the dashed lines -- at which there is a boundary between regions where only one strand crosses the $y$-$z$ plane and those where multiple strands do. Regions in which there are multiple crossers are identified, these are indicated by the shaded areas. They may be closed and the Alexander polynomial calculated to identify which of them contains the knot. (b) - (d) Subsequently, a finer determination of the knot position may be achieved by taking the knot-containing section and considering subsections of it. These are closed by extending the polymer in the $x$-direction, as shown by the dotted lines. The Alexander polynomial may then be calculated for each of these. The section with the correct Alexander polynomial that contains the least number of beads is taken as containing the knot. (b) - (d) show a few examples of subsections. The subsection shown in (c) would be identified: that in (b) is contains more beads and that in (d) has the wrong polynomial.}
\end{center}
\end{figure}

For simulations considering the positional probability distribution or size of the knot, it is necessary to determine the knotted section of the polymer. We applied a method, summarized in~\ref{fig:knot_find_var_flex_tens}, based on calculating the Alexander polynomial,~\cite{livingston} $A_k(x)$, at $x = -2$ for polymer subsections~\cite{katritch}. Since the polymer is extended in the $x$-direction by the tension, there will usually be $x$-positions at which only one part of polymer crosses the $y$-$z$-plane. Regions that are bounded by such points are considered. Only one will have the correct $A_k(-2)$. The more exact position is then found by taking subsections of this region, closing them with extensions in the $\pm x$-direction, and finding the shortest with the correct $A_k(-2)$. The center of this section is taken as the knot position and the number of beads it contains as the knot size. This is the same method applied for the determination of $N_{knot}$ for flexible chains earlier in this paper.

Our procedure may occasionally result in a false identification of a knot due to extra crossings included by the closing sections. However, in previous studies the rate of such errors was found to be low and to usually involve sections larger than truly knotted ones~\cite{katritch}. We thus do not expect such pitfalls to significantly affect our results but we refer the interested reader to an in-depth consideration of such schemes.~\cite{tubiana2} We also found that, occasionally, no $x$-positions with only one crossing of the $y$-$z$-plane were found. In this case, the knot position was not identified and so these configurations were neglected. The rate of such configurations was $< 1\%$ for all the results presented. As a further check we verified that, for the knot size results, if instead of neglecting the configurations, a knot size equal to the total polymer size was added, the final averages were not changed by more than the errorbars. Simulations with knot-finding were performed with the same MC algorithm as for the free energy calculations. However, due to the computational cost of the knot-finding algorithm, which would be difficult to implement on a GPU, the calculations were performed entirely on a CPU. 

\begin{figure}[htb!]
\begin{center}
\includegraphics[scale=0.55]{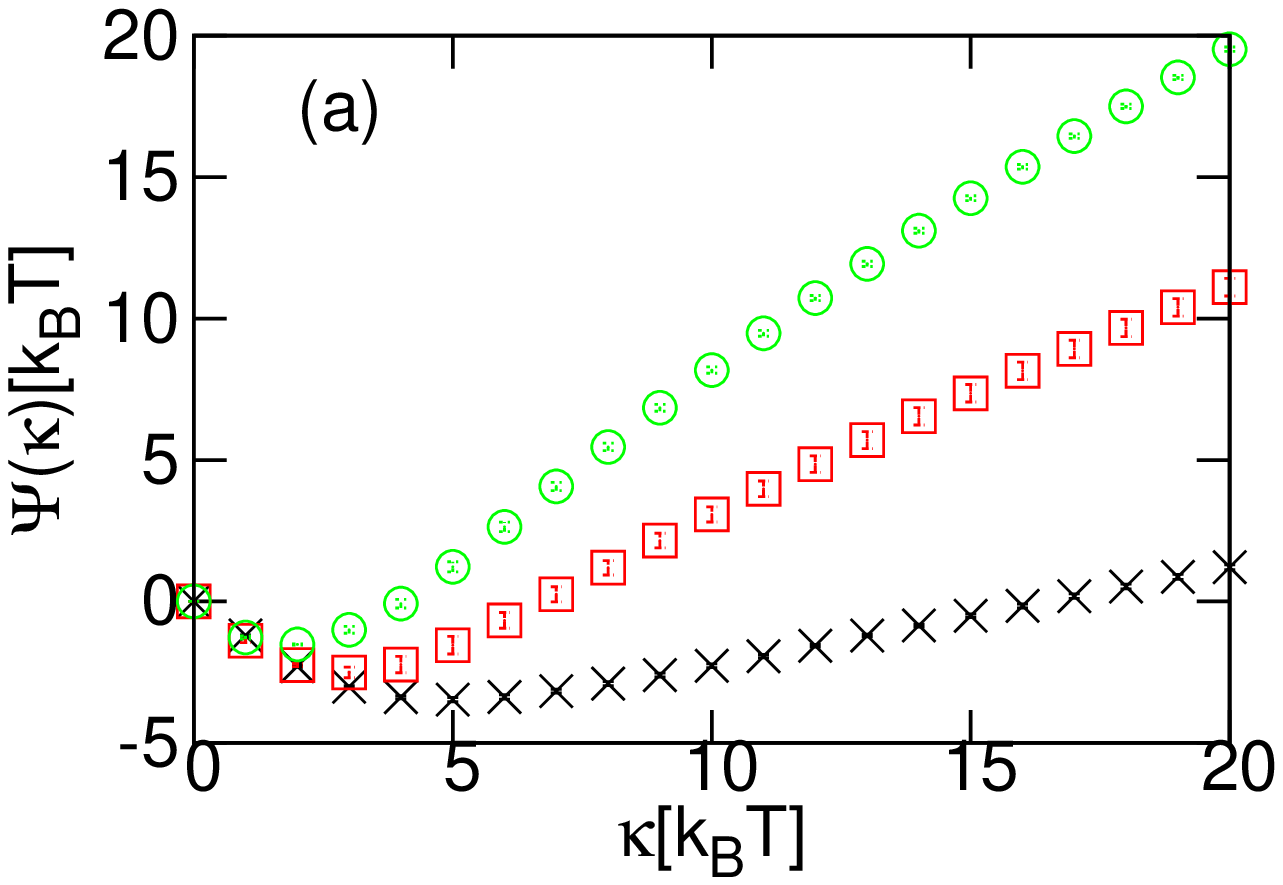}
\includegraphics[scale=0.55]{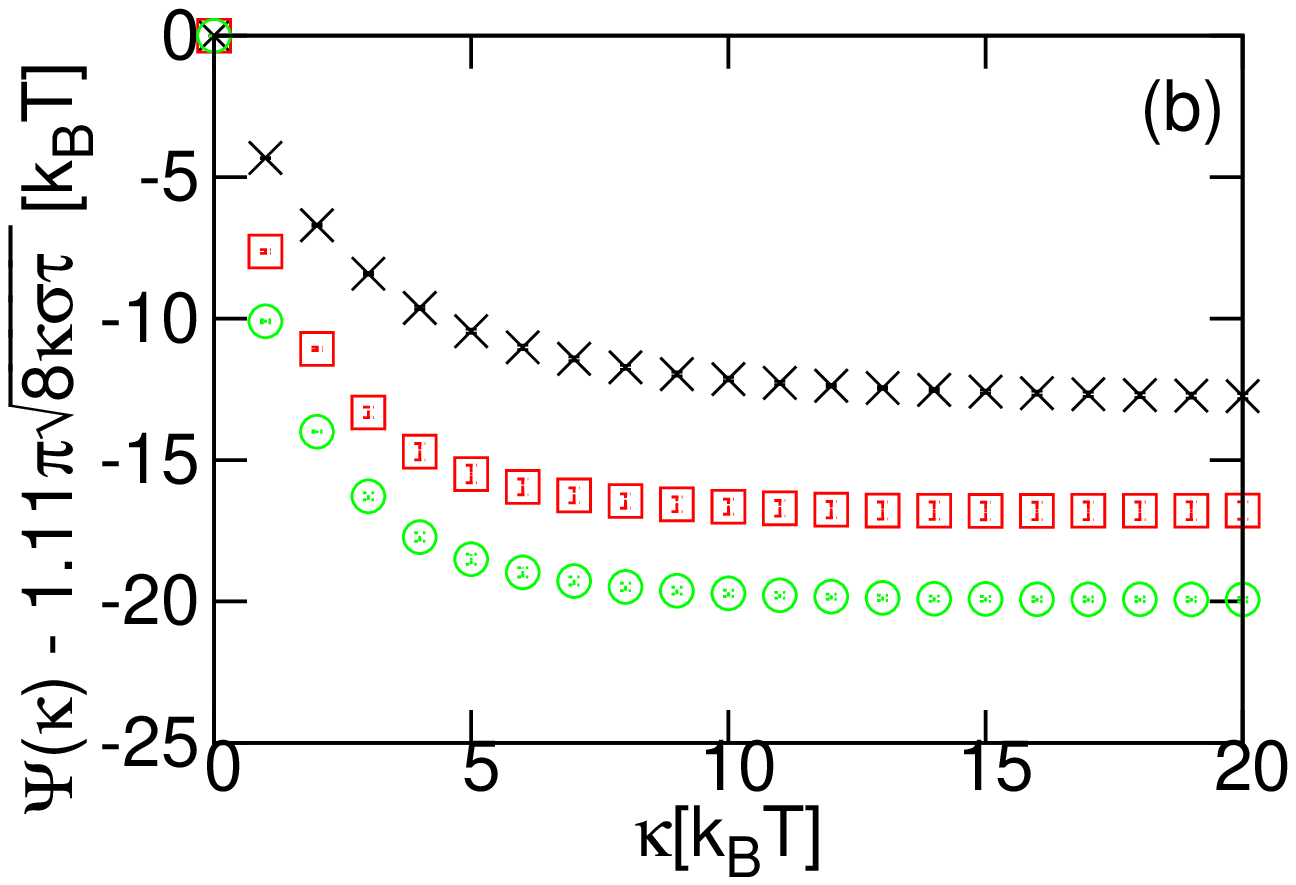}
\caption{\label{fig:delta_F_kap}  (a) The difference in free-energy, $\Psi(\kappa) \equiv \Delta F_{knotting}(\kappa) - \Delta F_{knotting}(0)$, against $\kappa$ for different tensions, $\tau$: $0.1 k_B T/\sigma$ ($\times$, black);  $0.4 k_B T/\sigma$ ($\square$, red);  $0.8 k_B T/\sigma$ ($\bigcirc$, green). Note the minimum at $\kappa = \kappa_{min}$, which decreases for increasing $\tau$. (b) The free-energy difference with a term proportional to the high $A$ limit in~\ref{eq:D_F_stiff} subtracted:  $\Psi(\kappa) - 1.11\sqrt{8 \pi^2 \kappa \sigma \tau}$ plotted against $\kappa$ for the same $\tau$.  Error bars were estimated by performing three independent repeats of the simulations.}
\end{center}
\end{figure}

We first present, in~\ref{fig:delta_F_kap}(a), results for $\Psi(\kappa)\equiv\Delta F_{knotting}(\kappa) - \Delta F_{knotting}(0)$ as a function of $\kappa$ for $\tau = 0.1$, $0.4$ and $0.8 k_BT/\sigma$. As expected, we observe that there is a minimum at non-zero $\kappa$, which we denote $\kappa_{min}$, and which decreases with increasing tension. In ~\ref{fig:delta_F_kap}(b) we also plot the same data subtracting a term proportional to $\sqrt{8 \pi^2 \kappa \sigma \tau}$, the expression for $\Delta F_{knotting}$ at high $A$ (\ref{eq:D_F_stiff} with $A = \sigma \kappa$). The additional proportionality factor of 1.11 was determined by fitting $\Delta F_{knotting}(\kappa) - \Delta F_{knotting}(0)$ for $\tau = 0.4$ and $0.8 k_BT/\sigma$ for $\kappa \ge 15 k_BT$. For both, the same factor was found to the accuracy that is given. The extra factor is likely necessary because our polymers do not have negligible thickness. To within errors, the curves for  $\tau = 0.4$ and $0.8  k_BT/\sigma$, with the expression subtracted, become flat for higher $\kappa$. This suggests that for these $\kappa$ values we have reached the regime where $\Delta F_{knotting}$ is dominated by the bending and tension terms. We further observe that, at the position of the minimum of the knotting free energy cost, the quantity $\Delta F_{knotting}(\kappa) - \Delta F_{knotting}(0) - 1.11\sqrt{8 \pi^2 \kappa \sigma \tau}$ still has a relatively steep slope, confirming that the entropic contribution is important in determining the position of the minimum.

\begin{figure}[htb!]
\begin{center}
\includegraphics[scale=0.55]{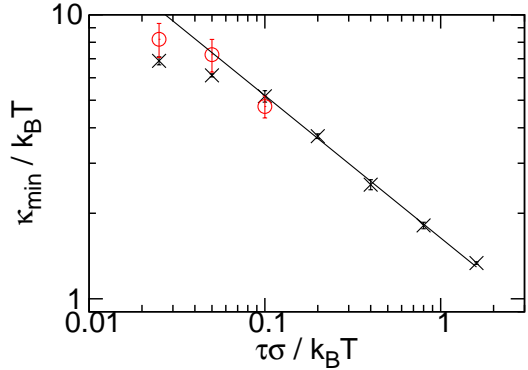}
\caption{\label{fig:kap_min_tens}  The minimum value $\kappa_{min}$ of $\Delta F_{knotting}$ against the applied tension $\tau$ for $N = 256$ ($\times$, black) and $N = 512$ ($\bigcirc$, red). The solid line is a fit to the five data points for $N = 256$ with highest $\tau$ values, it has a slope of $-0.50 \pm 0.01$. Errorbars were estimated by performing three independent repeats of the simulations.}
\end{center}
\end{figure}

In~\ref{fig:kap_min_tens}, we show the dependence of $\kappa_{min}$ on $\tau$ for $N = 256$. Plotting on a logarithmic scale, we see that the points for the highest five $\tau$ show a power-law relationship. Fitting to these data, we find an exponent of $-0.50 \pm 0.01$. We thus obtain a power-law dependence of the optimal rigidity on the tension that we anticipated in~\ref{eq:scalea}, but with an exponent different than the $t = -0.43$ we found from~\ref{fig:ks_tens}, as expected. For the lowest two $\tau$ we see that the curve deviates from this power-law relationship. This may be attributed to finite size effects. To verify this we repeated simulations for the three lowest $\tau$ for $N = 512$: the results are also plotted in~\ref{fig:kap_min_tens}. We observe that, as expected, the results are consistent with the same power-law relationship and also follow it to lower $\tau$. 

\begin{figure}[htb!]
\begin{center}
\includegraphics[scale=0.55]{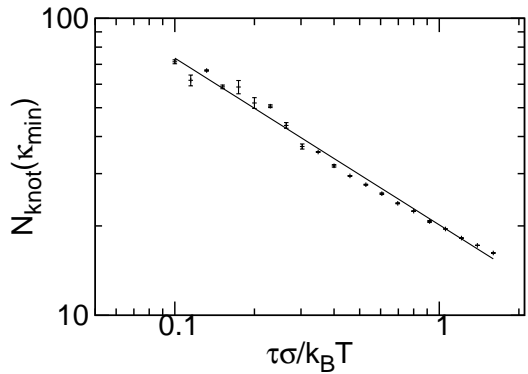}
\caption{\label{fig:ks_kap_min_tens}  The number of bead in the knot at $\kappa_{min}$, $N_{knot}(\kappa_{min})$ against $\tau$. The solid line is a fit with a slope of $-0.56 \pm 0.02$. Error bars were estimated by performing two independent repeats of the simulations.}
\end{center}
\end{figure}

We expect $\kappa_{min}$ to be approximately that value of bending rigidity for which the size of the knot is equal to the bending length. We consider the variation of the number of the beads in the knot at $\kappa_{min}$, $N_{knot}(\kappa_{min})$, with $\tau$. We take $\kappa_{min}$ to be given by the best fit relationship from~\ref{fig:kap_min_tens}. We plot the results for  $N_{knot}(\kappa_{min})$ in~\ref{fig:ks_kap_min_tens}. By fitting, we find an exponent of $-0.56\pm0.02$, close to $-0.50 \pm 0.01$: indeed, $N_{knot}(\kappa_{min}) \sim \kappa_{min}$ because the polymer is stiff at the scale of the knot.

\begin{figure}[htb!]
\begin{center}
\includegraphics[scale=0.55]{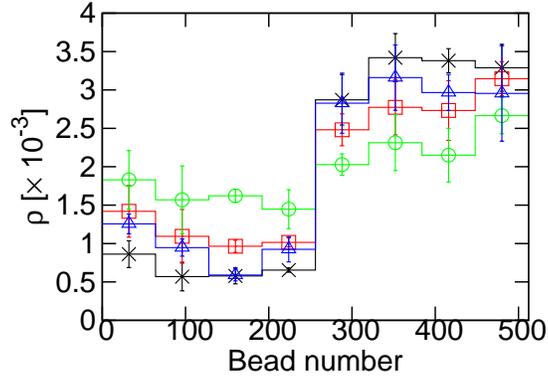}
\caption{\label{fig:prob_dens_pos}  Probability density, $\rho$, of finding the knot at a given position along the polymer under tension, $\tau = 0.8 k_BT/\sigma$. For beads $256 - 511$, $\kappa_{i} = 1.806 k_BT \approx \kappa_{min}$, whilst for beads $0-255$ $\kappa_{i} =  0$ ($\times$, black), $\kappa_{i} = 0.4353 k_BT$ ($\square$ ,red), $\kappa_{i} = 0.8706 k_BT$ ($\bigcirc$, green) or $\kappa_{i} = 3.842 k_BT$ ($\triangle$, blue). Errorbars were estimated by performing three independent repeats of the simulations.}
\end{center}
\end{figure}

We have found that $\Delta F_{knotting}$ has a minimum at a non-zero value of the bending stiffness, namely  $\kappa_{min}$. We therefore expect that, if we consider a knotted polymer with non-uniform flexibility under tension, $\tau$, the knot will be more likely to be found in a region with $\kappa_{min}$ than in other regions. To test this, we consider a polymer of $N = 512$  beads at $\tau = 0.8 k_BT/\sigma$, split into two halves: the first 256 beads have $\kappa_{i} = \kappa_{0} \neq \kappa_{min}$. The second 256 beads have $\kappa_{i} = 1.806 k_BT \approx \kappa_{min}$ for this $\tau$. In~\ref{fig:prob_dens_pos}, we plot results for $\kappa_{0} = 0$, $0.4353 k_BT$,  $0.8706 k_BT$ and $3.842 k_BT$, i.e. three regions with $\kappa_0 < \kappa_{min}$ and one with $\kappa_0 > \kappa_{min}$. Results are binned into 8 bins of 64 beads each. In each case we find that the probability of finding the knot in the region with $\kappa_{min}$ is higher. In other words, the knot prefers to localize in the region where $\kappa \approx \kappa_0$. Furthermore, we find that the probabilities are approximately those that would be expected from the free energy calculations. For $\kappa_0 = 0$ in~\ref{fig:prob_dens_pos}, the ratio between the average of the first four bins and that of the second four is $4.9\pm0.5$, giving an expected free energy difference of $1.6\pm0.1 k_BT$. The minimum $\Delta F_{knotting}(\kappa) - \Delta F_{knotting}(0)$ for $\tau = 0.8 k_BT/\sigma$ in~\ref{fig:delta_F_kap}(a) is $-1.52\pm0.02 k_BT$.
 
To summarize, inspired by correlations between polymer flexibility and knotting seen in biology, we have investigated how the cost of forming a knot in a polymer under tension, $\tau$, depends on the polymer's stiffness, controlled in our model by $\kappa$. For high $\kappa$, our results agree with a simple expression including only bending and tension, whilst for lower $\kappa$ entropy must also be taken into account. There is a non-zero minimum of the free energy difference between unknotted and knotted states at $\kappa =\kappa_{min}$. The position of the minimum is seen to depend on tension as $\kappa_{min} \sim \tau^{-0.5}$. We argue that $\kappa_{min}$ is determined by the relative sizes of the knot and the bending length and find that the number of polymer beads in the knot at $\kappa_{min}$ is consistent with this argument. We considered knotted polymers with two sections with different $\kappa$ and found that the knot is more likely to be found in the section with $\kappa_{min}$.

Biological DNA is typically highly confined and in future work it would be interesting to investigate the effect of confinement on the results we have observed~\cite{dai,matthews2011}. It would also be interesting to investigate how the position of cleavage sites relative to regions of different flexibility affects the steady state level of knotting~\cite{volo2}, as well as looking into how the effect of flexibility may combine with previously suggested topoisomerase II guidance mechanisms.~\cite{liu3} Finally, it would be intriguing to investigate how non-uniform flexibility affects the diffusional dynamics of a knot along a polymer~\cite{huang,matthews2010}.

\begin{acknowledgement}

The computational results presented have been achieved using the Vienna Scientific Cluster (VSC). This work was supported by the Austrian Science Fund (FWF): P 23400-N16 and M1367, and by the EPSRC through a DTA studentship to RM. We thank J.M. Yeomans for helpful discussions.

\end{acknowledgement}



\bibliography{bend_rigid_knot.bib}

\providecommand*\mcitethebibliography{\thebibliography}
\csname @ifundefined\endcsname{endmcitethebibliography}
  {\let\endmcitethebibliography\endthebibliography}{}
\begin{mcitethebibliography}{33}
\providecommand*\natexlab[1]{#1}
\providecommand*\mciteSetBstSublistMode[1]{}
\providecommand*\mciteSetBstMaxWidthForm[2]{}
\providecommand*\mciteBstWouldAddEndPuncttrue
  {\def\EndOfBibitem{\unskip.}}
\providecommand*\mciteBstWouldAddEndPunctfalse
  {\let\EndOfBibitem\relax}
\providecommand*\mciteSetBstMidEndSepPunct[3]{}
\providecommand*\mciteSetBstSublistLabelBeginEnd[3]{}
\providecommand*\EndOfBibitem{}
\mciteSetBstSublistMode{f}
\mciteSetBstMaxWidthForm{subitem}{(\alph{mcitesubitemcount})}
\mciteSetBstSublistLabelBeginEnd
  {\mcitemaxwidthsubitemform\space}
  {\relax}
  {\relax}

\bibitem[Rybenkov et~al.(1997)Rybenkov, Ullsperger, Vologodskii, and
  Cozzarelli]{rybenkov}
Rybenkov,~V.; Ullsperger,~C.; Vologodskii,~A.; Cozzarelli,~N. \emph{Science}
  \textbf{1997}, \emph{277}, 690--693\relax
\mciteBstWouldAddEndPuncttrue
\mciteSetBstMidEndSepPunct{\mcitedefaultmidpunct}
{\mcitedefaultendpunct}{\mcitedefaultseppunct}\relax
\EndOfBibitem
\bibitem[Burden and Osheroff(1999)Burden, and Osheroff]{burden}
Burden,~D.; Osheroff,~N. \emph{J. Biol. Chem.} \textbf{1999}, \emph{274},
  5227--5235\relax
\mciteBstWouldAddEndPuncttrue
\mciteSetBstMidEndSepPunct{\mcitedefaultmidpunct}
{\mcitedefaultendpunct}{\mcitedefaultseppunct}\relax
\EndOfBibitem
\bibitem[Adachi et~al.(1989)Adachi, K{\"a}s, and Laemmli]{adachi}
Adachi,~Y.; K{\"a}s,~E.; Laemmli,~U. \emph{EMBO J.} \textbf{1989}, \emph{8},
  3997--4006\relax
\mciteBstWouldAddEndPuncttrue
\mciteSetBstMidEndSepPunct{\mcitedefaultmidpunct}
{\mcitedefaultendpunct}{\mcitedefaultseppunct}\relax
\EndOfBibitem
\bibitem[Razin et~al.(1991)Razin, Vassetzky, and Hancock]{razin}
Razin,~S.; Vassetzky,~Y.; Hancock,~R. \emph{Biochem. Biophys. Res. Commun.}
  \textbf{1991}, \emph{177}, 265--270\relax
\mciteBstWouldAddEndPuncttrue
\mciteSetBstMidEndSepPunct{\mcitedefaultmidpunct}
{\mcitedefaultendpunct}{\mcitedefaultseppunct}\relax
\EndOfBibitem
\bibitem[Masliah et~al.(2008)Masliah, Ren{\'e}, Zargarian, Fermandjian, and
  Mauffret]{masliah}
Masliah,~G.; Ren{\'e},~B.; Zargarian,~L.; Fermandjian,~S.; Mauffret,~O.
  \emph{J. Mol. Biol.} \textbf{2008}, \emph{381}, 692--706\relax
\mciteBstWouldAddEndPuncttrue
\mciteSetBstMidEndSepPunct{\mcitedefaultmidpunct}
{\mcitedefaultendpunct}{\mcitedefaultseppunct}\relax
\EndOfBibitem
\bibitem[Hogan et~al.(1983)Hogan, LeGrange, and Austin]{hogan}
Hogan,~M.; LeGrange,~J.; Austin,~B. \emph{Nature} \textbf{1983}, \emph{304},
  752--754\relax
\mciteBstWouldAddEndPuncttrue
\mciteSetBstMidEndSepPunct{\mcitedefaultmidpunct}
{\mcitedefaultendpunct}{\mcitedefaultseppunct}\relax
\EndOfBibitem
\bibitem[Okonogi et~al.(2002)Okonogi, Alley, Reese, Hopkins, and
  Robinson]{okonogi}
Okonogi,~T.; Alley,~S.; Reese,~A.; Hopkins,~P.; Robinson,~B. \emph{Biophys. J.}
  \textbf{2002}, \emph{83}, 3446--3459\relax
\mciteBstWouldAddEndPuncttrue
\mciteSetBstMidEndSepPunct{\mcitedefaultmidpunct}
{\mcitedefaultendpunct}{\mcitedefaultseppunct}\relax
\EndOfBibitem
\bibitem[Scipioni et~al.(2002)Scipioni, Anselmi, Zuccheri, Samori, and
  De~Santis]{scipioni}
Scipioni,~A.; Anselmi,~C.; Zuccheri,~G.; Samori,~B.; De~Santis,~P.
  \emph{Biophys. J.} \textbf{2002}, \emph{83}, 2408--2418\relax
\mciteBstWouldAddEndPuncttrue
\mciteSetBstMidEndSepPunct{\mcitedefaultmidpunct}
{\mcitedefaultendpunct}{\mcitedefaultseppunct}\relax
\EndOfBibitem
\bibitem[Ramstein and Lavery(1988)Ramstein, and Lavery]{ramstein}
Ramstein,~J.; Lavery,~R. \emph{Proc. Nat. Acad. Sci.} \textbf{1988}, \emph{85},
  7231--7235\relax
\mciteBstWouldAddEndPuncttrue
\mciteSetBstMidEndSepPunct{\mcitedefaultmidpunct}
{\mcitedefaultendpunct}{\mcitedefaultseppunct}\relax
\EndOfBibitem
\bibitem[Dong and Berger(2007)Dong, and Berger]{dong}
Dong,~K.; Berger,~J. \emph{Nature} \textbf{2007}, \emph{450}, 1201--1205\relax
\mciteBstWouldAddEndPuncttrue
\mciteSetBstMidEndSepPunct{\mcitedefaultmidpunct}
{\mcitedefaultendpunct}{\mcitedefaultseppunct}\relax
\EndOfBibitem
\bibitem[Livingston(1993)]{livingston}
Livingston,~C. \emph{Knot Theory}; The Mathematical Association of America:
  Washington, 1993\relax
\mciteBstWouldAddEndPuncttrue
\mciteSetBstMidEndSepPunct{\mcitedefaultmidpunct}
{\mcitedefaultendpunct}{\mcitedefaultseppunct}\relax
\EndOfBibitem
\bibitem[Liu et~al.(2009)Liu, Deibler, Chan, and Zechiedrich]{liu3}
Liu,~Z.; Deibler,~R.; Chan,~H.; Zechiedrich,~L. \emph{Nucl. Acid Res.}
  \textbf{2009}, \emph{37}, 661--671\relax
\mciteBstWouldAddEndPuncttrue
\mciteSetBstMidEndSepPunct{\mcitedefaultmidpunct}
{\mcitedefaultendpunct}{\mcitedefaultseppunct}\relax
\EndOfBibitem
\bibitem[Bustamante et~al.(2003)Bustamante, Bryant, and Smith]{bustamante2003}
Bustamante,~C.; Bryant,~Z.; Smith,~S. \emph{Nature} \textbf{2003}, \emph{421},
  423--426\relax
\mciteBstWouldAddEndPuncttrue
\mciteSetBstMidEndSepPunct{\mcitedefaultmidpunct}
{\mcitedefaultendpunct}{\mcitedefaultseppunct}\relax
\EndOfBibitem
\bibitem[Farago et~al.(2002)Farago, Kantor, and Kardar]{farago}
Farago,~O.; Kantor,~Y.; Kardar,~M. \emph{Europhys. Lett.} \textbf{2002},
  \emph{60}, 53--59\relax
\mciteBstWouldAddEndPuncttrue
\mciteSetBstMidEndSepPunct{\mcitedefaultmidpunct}
{\mcitedefaultendpunct}{\mcitedefaultseppunct}\relax
\EndOfBibitem
\bibitem[Marcone et~al.(2005)Marcone, Orlandini, Stella, and Zonta]{marcone}
Marcone,~B.; Orlandini,~E.; Stella,~A.; Zonta,~F. \emph{J. Phys. A: Math. Gen.}
  \textbf{2005}, \emph{38}, L15--L21\relax
\mciteBstWouldAddEndPuncttrue
\mciteSetBstMidEndSepPunct{\mcitedefaultmidpunct}
{\mcitedefaultendpunct}{\mcitedefaultseppunct}\relax
\EndOfBibitem
\bibitem[Mansfield and Douglas(2010)Mansfield, and Douglas]{mansfield2010}
Mansfield,~M.; Douglas,~J. \emph{J. Chem. Phys.} \textbf{2010}, \emph{133},
  044903\relax
\mciteBstWouldAddEndPuncttrue
\mciteSetBstMidEndSepPunct{\mcitedefaultmidpunct}
{\mcitedefaultendpunct}{\mcitedefaultseppunct}\relax
\EndOfBibitem
\bibitem[Ercolini et~al.(2007)Ercolini, Valle, Adamcik, Witz, Metzler,
  De~Los~Rios, Roca, and Dietler]{ercolini}
Ercolini,~E.; Valle,~F.; Adamcik,~J.; Witz,~G.; Metzler,~R.; De~Los~Rios,~P.;
  Roca,~J.; Dietler,~G. \emph{Phys. Rev. Lett.} \textbf{2007}, \emph{98},
  058102\relax
\mciteBstWouldAddEndPuncttrue
\mciteSetBstMidEndSepPunct{\mcitedefaultmidpunct}
{\mcitedefaultendpunct}{\mcitedefaultseppunct}\relax
\EndOfBibitem
\bibitem[Tubiana et~al.(2011)Tubiana, Orlandini, and Micheletti]{tubiana}
Tubiana,~L.; Orlandini,~E.; Micheletti,~C. \emph{Phys. Rev. Lett.}
  \textbf{2011}, \emph{107}, 188302\relax
\mciteBstWouldAddEndPuncttrue
\mciteSetBstMidEndSepPunct{\mcitedefaultmidpunct}
{\mcitedefaultendpunct}{\mcitedefaultseppunct}\relax
\EndOfBibitem
\bibitem[Zheng and Vologodskii(2010)Zheng, and Vologodskii]{zheng2010}
Zheng,~X.; Vologodskii,~A. \emph{Phys. Rev. E} \textbf{2010}, \emph{81},
  041806\relax
\mciteBstWouldAddEndPuncttrue
\mciteSetBstMidEndSepPunct{\mcitedefaultmidpunct}
{\mcitedefaultendpunct}{\mcitedefaultseppunct}\relax
\EndOfBibitem
\bibitem[de~Gennes(1979)]{deGennes_scpp}
de~Gennes,~P. \emph{{Scaling Concepts in Polymer Physics}}; Cornell University
  Press: New York, 1979\relax
\mciteBstWouldAddEndPuncttrue
\mciteSetBstMidEndSepPunct{\mcitedefaultmidpunct}
{\mcitedefaultendpunct}{\mcitedefaultseppunct}\relax
\EndOfBibitem
\bibitem[Gallotti and Pierre-Louis(2007)Gallotti, and Pierre-Louis]{gallotti}
Gallotti,~R.; Pierre-Louis,~O. \emph{Phys. Rev. E} \textbf{2007}, \emph{75},
  031801\relax
\mciteBstWouldAddEndPuncttrue
\mciteSetBstMidEndSepPunct{\mcitedefaultmidpunct}
{\mcitedefaultendpunct}{\mcitedefaultseppunct}\relax
\EndOfBibitem
\bibitem[Chen and Sullivan(2006)Chen, and Sullivan]{chen2006}
Chen,~J.; Sullivan,~D. \emph{Macromolecules.} \textbf{2006}, \emph{39},
  7769--7773\relax
\mciteBstWouldAddEndPuncttrue
\mciteSetBstMidEndSepPunct{\mcitedefaultmidpunct}
{\mcitedefaultendpunct}{\mcitedefaultseppunct}\relax
\EndOfBibitem
\bibitem[Frenkel and Smit(2002)Frenkel, and Smit]{frenkel}
Frenkel,~D.; Smit,~B. \emph{{Understanding Molecular Simulation: from
  Algorithms to Applications}}; Academic Press: London, 2002\relax
\mciteBstWouldAddEndPuncttrue
\mciteSetBstMidEndSepPunct{\mcitedefaultmidpunct}
{\mcitedefaultendpunct}{\mcitedefaultseppunct}\relax
\EndOfBibitem
\bibitem[Mehlig et~al.(1992)Mehlig, Heermann, and Forrest]{mehlig}
Mehlig,~B.; Heermann,~D.; Forrest,~B. \emph{Phys. Rev. B} \textbf{1992},
  \emph{45}, 679--685\relax
\mciteBstWouldAddEndPuncttrue
\mciteSetBstMidEndSepPunct{\mcitedefaultmidpunct}
{\mcitedefaultendpunct}{\mcitedefaultseppunct}\relax
\EndOfBibitem
\bibitem[van Meel et~al.(2008)van Meel, Arnold, Frenkel, Zwart, and
  Belleman]{vanMeel2008}
van Meel,~J.; Arnold,~A.; Frenkel,~D.; Zwart,~S.; Belleman,~R. \emph{Molecular
  Simulation} \textbf{2008}, \emph{34}, 259--266\relax
\mciteBstWouldAddEndPuncttrue
\mciteSetBstMidEndSepPunct{\mcitedefaultmidpunct}
{\mcitedefaultendpunct}{\mcitedefaultseppunct}\relax
\EndOfBibitem
\bibitem[Katritch et~al.(2000)Katritch, Olson, Vologodskii, Dubochet, and
  Stasiak]{katritch}
Katritch,~V.; Olson,~W.; Vologodskii,~A.; Dubochet,~J.; Stasiak,~A. \emph{Phys.
  Rev. E} \textbf{2000}, \emph{61}, 5545--5549\relax
\mciteBstWouldAddEndPuncttrue
\mciteSetBstMidEndSepPunct{\mcitedefaultmidpunct}
{\mcitedefaultendpunct}{\mcitedefaultseppunct}\relax
\EndOfBibitem
\bibitem[Tubiana et~al.(2011)Tubiana, Orlandini, and Micheletti]{tubiana2}
Tubiana,~L.; Orlandini,~E.; Micheletti,~C. \emph{Prog. Theor. Phys. Suppl.}
  \textbf{2011}, \emph{191}, 192--204\relax
\mciteBstWouldAddEndPuncttrue
\mciteSetBstMidEndSepPunct{\mcitedefaultmidpunct}
{\mcitedefaultendpunct}{\mcitedefaultseppunct}\relax
\EndOfBibitem
\bibitem[Dai et~al.(2012)Dai, van~der Maarel, and Doyle]{dai}
Dai,~L.; van~der Maarel,~J.; Doyle,~P. \emph{ACS Macro Letters} \textbf{2012},
  \emph{1}, 732--736\relax
\mciteBstWouldAddEndPuncttrue
\mciteSetBstMidEndSepPunct{\mcitedefaultmidpunct}
{\mcitedefaultendpunct}{\mcitedefaultseppunct}\relax
\EndOfBibitem
\bibitem[Matthews et~al.(2011)Matthews, Louis, and Yeomans]{matthews2011}
Matthews,~R.; Louis,~A.; Yeomans,~J. \emph{Mol. Phys.} \textbf{2011},
  \emph{109}, 1289--1295\relax
\mciteBstWouldAddEndPuncttrue
\mciteSetBstMidEndSepPunct{\mcitedefaultmidpunct}
{\mcitedefaultendpunct}{\mcitedefaultseppunct}\relax
\EndOfBibitem
\bibitem[Vologodskii et~al.(2001)Vologodskii, Zhang, Rybenkov, Podtelezhnikov,
  Subramanian, Griffith, and Cozzarelli]{volo2}
Vologodskii,~A.; Zhang,~W.; Rybenkov,~V.; Podtelezhnikov,~A.; Subramanian,~D.;
  Griffith,~J.; Cozzarelli,~N. \emph{Proc. Nat. Acad. Sci.} \textbf{2001},
  \emph{98}, 3045--3049\relax
\mciteBstWouldAddEndPuncttrue
\mciteSetBstMidEndSepPunct{\mcitedefaultmidpunct}
{\mcitedefaultendpunct}{\mcitedefaultseppunct}\relax
\EndOfBibitem
\bibitem[Huang and Makarov(2007)Huang, and Makarov]{huang}
Huang,~L.; Makarov,~D. \emph{J. Phys. Chem. A} \textbf{2007}, \emph{111},
  10338--10344\relax
\mciteBstWouldAddEndPuncttrue
\mciteSetBstMidEndSepPunct{\mcitedefaultmidpunct}
{\mcitedefaultendpunct}{\mcitedefaultseppunct}\relax
\EndOfBibitem
\bibitem[Matthews et~al.(2010)Matthews, Louis, and Yeomans]{matthews2010}
Matthews,~R.; Louis,~A.; Yeomans,~J. \emph{EPL} \textbf{2010}, \emph{89},
  20001\relax
\mciteBstWouldAddEndPuncttrue
\mciteSetBstMidEndSepPunct{\mcitedefaultmidpunct}
{\mcitedefaultendpunct}{\mcitedefaultseppunct}\relax
\EndOfBibitem
\end{mcitethebibliography}

%% graphical table of contents.
%%%%%%%%%%%%%%%%%%%%%%%%%%%%%%%%%%%%%%%%%%%%%%%%%%%%%%%%%%%%%%%%%%%%%

\begin{tocentry}

\includegraphics[scale=0.208]{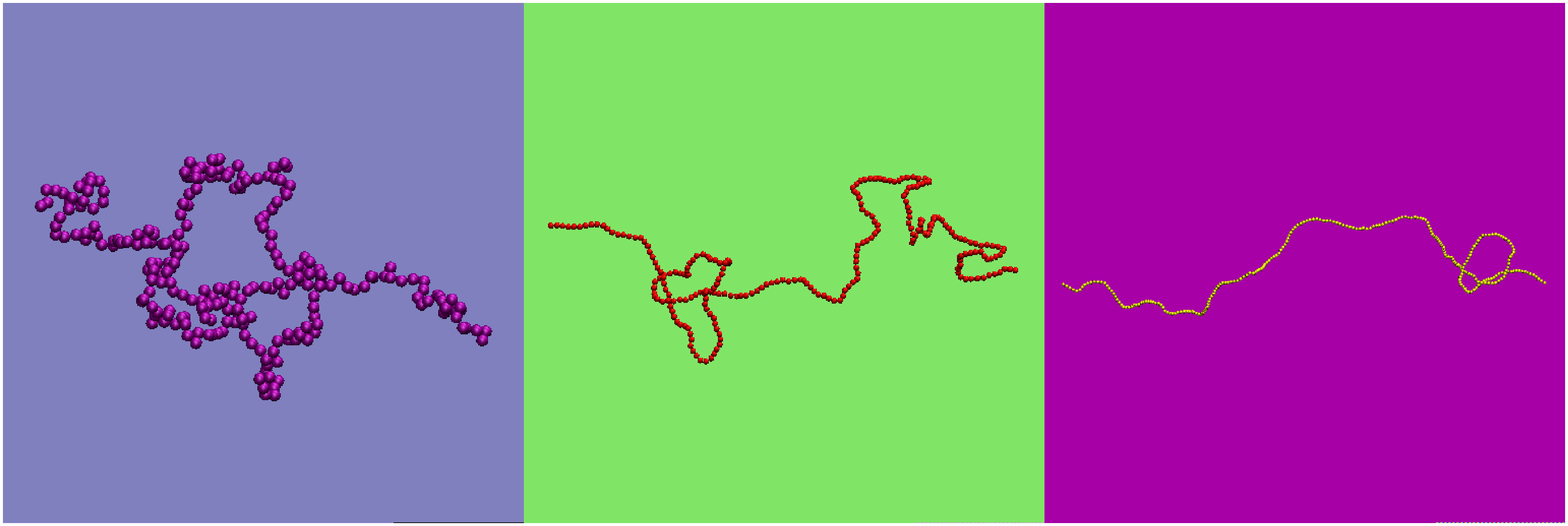}

\end{tocentry}

\end{document}